\documentclass[aps,pra,twocolumn,groupedaddress]{revtex4}
\usepackage{graphicx}
\usepackage{multirow}
\usepackage{amsmath}
\usepackage{amssymb}
\usepackage[latin1]{inputenc}
\usepackage{array}
\usepackage{color}
\usepackage[caption=false]{subfig}

\begin{document}

\title{Noise spectroscopy with large clouds of cold atoms}
\author{Samir Vartabi Kashanian$^{1,2}$}
\author{Aur\'elien Eloy$^{1}$}
\author{William Guerin$^1$}
\author{Michel Lintz$^2$}
\author{Mathilde Fouch\'e$^1$} \email{mathilde.fouche@inln.cnrs.fr}
\author{Robin Kaiser$^1$}

\affiliation{$^{1}$Universit\'e C\^ote d'Azur, CNRS, INLN, France}
\affiliation{$^{2}$Universit\'e C\^ote d'Azur, OCA, CNRS, ARTEMIS, France}

\date{\today}

\begin{abstract}
Noise measurement is a powerful tool to investigate many phenomena from laser characterization to quantum behavior of light. In this paper, we report on intensity noise measurements obtained when a laser beam is transmitted through a large cloud of cold atoms. While this measurement could possibly investigate complex processes such as the influence of atomic motion, one is first limited by the conversion of the intrinsic laser frequency noise to intensity noise via the atomic resonance. This conversion is studied here in details. We show that, while experimental intensity noise spectra collapse onto the same curve at low Fourier frequencies, some differences appear at higher frequencies when the probe beam is detuned from the center of the resonance line. A simple model, based on a mean-field approach, which corresponds to describing the atomic cloud by a dielectric susceptibility, is sufficient to understand the main features. Using this model, the noise spectra allow extracting some quantitative informations on the laser noise as well as on the atomic sample.
\end{abstract}


\maketitle

\section{Introduction}

Progress in the study of light-matter interaction has opened the way to important developments in quantum optics. In particular, laser-cooled atoms are used to develop quantum memories\,\cite{Sangouard_2011}, novel laser designs\,\cite{Guerin_2008, Bohnet_2012}, or to prepare non classical states for potential use in metrology\,\cite{Hosten_2016, Cox_2016}.
To investigate the nonclassical behavior of light, the study of average intensities is not sufficient and one needs to measure coincidence rates\,\cite{Clauser_1969,Aspect_1982}, correlation and anticorrelation functions\,\cite{Kimble_1977,Aspect_1982,URen_2005,Chaneliere_2005,Peyronel_2012} or noise and fluctuations\,\cite{Silberhorn_2001,Josse_2003,Marino_2008,Agha_10}.
Measuring the light fluctuations and correlations after their interaction with cold atomic samples also provides information on the atomic motion\,\cite{Jurczak_1995,Jurczak_1996,Grover_2015} and could be used to characterize more subtle effects due to interference effects in multiple scattering\,\cite{Mueller_2015} or obtain direct evidence of the random laser operation in cold atoms\,\cite{Baudouin_2013}.

In this work, we address a particular configuration in which intensity noise measurements are performed on a laser beam \emph{transmitted} through a sample of laser-cooled atoms. This transmission geometry is relevant to investigate different properties, such as the reduction of the noise below the shot-noise level (squeezing)\,\cite{Lambrecht_1996,Ries_2003}, the extra noise due to the atomic internal structure via Raman scattering\,\cite{Lezama_2015}, the cooperative fluorescence from a strongly driven dilute cloud of atoms\,\cite{Ott_2013}, or two-photon optical nonlinearity\,\cite{Peyronel_2012}. However in this kind of geometry, and contrary to the fluorescence configuration, the contribution of the intrinsic noise of the involved lasers can be especially important, with in particular the conversion of laser frequency or phase noise to intensity noise through the atomic resonance. This technical noise may be hard to distinguish from the signal under study and a good understanding of this process is thus essential.

The frequency to intensity conversion, in which the atomic resonance acts as a frequency discriminator, was first reported in\,\cite{Yabuzaki_1991}. It was then studied theoretically\,\cite{Walser_1994, Iyyanki_1995, Vasavada_1995} as well as experimentally, using atomic or molecular resonance, either to measure the laser properties \,\cite{Myers_2002, Bartalini_2010}, to extract atomic characteristics\,\cite{McLean_1993, Rosenbluh_1998} or to study the light-matter interaction\,\cite{McIntyre_1993, Camparo_1999, Bahoura_2001, Townsend_2005}. This phenomenon has been thus extensively studied. However, all previous experimental studies used room-temperature or hot vapors. In this case, the Doppler effect needs to be taken into account and limits the atomic spectral linewidth. On the other hand, cold atomic samples correspond to a system where the Doppler effect can generally be ignored. The atomic spectral linewidth is reduced, improving the frequency to intensity noise conversion. This results in a quantitative change of the noise spectra, but also in a qualitative change as we will see in this paper, with in particular the appearance of structures for Fourier frequencies higher than the atomic linewidth.

In this paper, we study the frequency to intensity noise conversion of a laser going through a \emph{cold} atomic cloud. The experimental setup and the results are presented in Sec.\,\ref{Sec:ColdAtoms}. Whereas all the noise spectra collapse onto the same curve at low Fourier frequencies, some differences appear at higher Fourier frequencies when the probe beam is detuned by various amonts from the center of the resonance line. We show in Sec.\,\ref{Sec:LowFreq} that the low Fourier frequencies components are well understood using the frequency discriminator approach, corresponding to what has been already obtained with room temperature or hot vapors. The apparent discrepancies at high Fourier frequencies are addressed in Sec.\,\ref{Sec:HighFreq}. We will show that a `mean-field approach'\,\cite{Javanainen_2016}, in which the atomic cloud is described by a complex index of refraction, is sufficient to explain, qualitatively and quantitatively, the measured spectra of the intensity noise, even when the on-resonance optical thickness is large. Finally, Appendix\,\ref{App:FP} presents the characterization of our probe laser with standard techniques, which serves as a benchmark for our measurements with cold atoms.

\section{Noise spectroscopy with cold atoms}\label{Sec:ColdAtoms}

\subsection{Apparatus} \label{subsec:apparatus}

\begin{figure}[h]
	\label{fig:setup}
	\centering
		 \subfloat[\label{fig:Cold_atoms_setup}]{\includegraphics[width=0.8\linewidth]{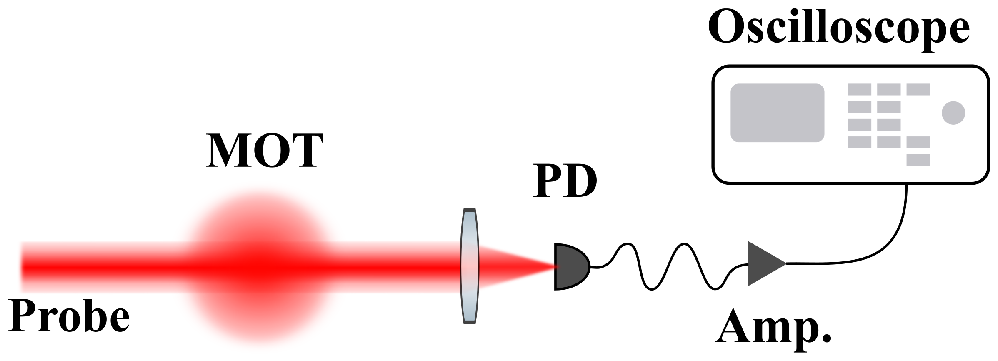}} \\
		 \subfloat[\label{fig:Cold_atoms_TS}]{\includegraphics[width=1.\linewidth]{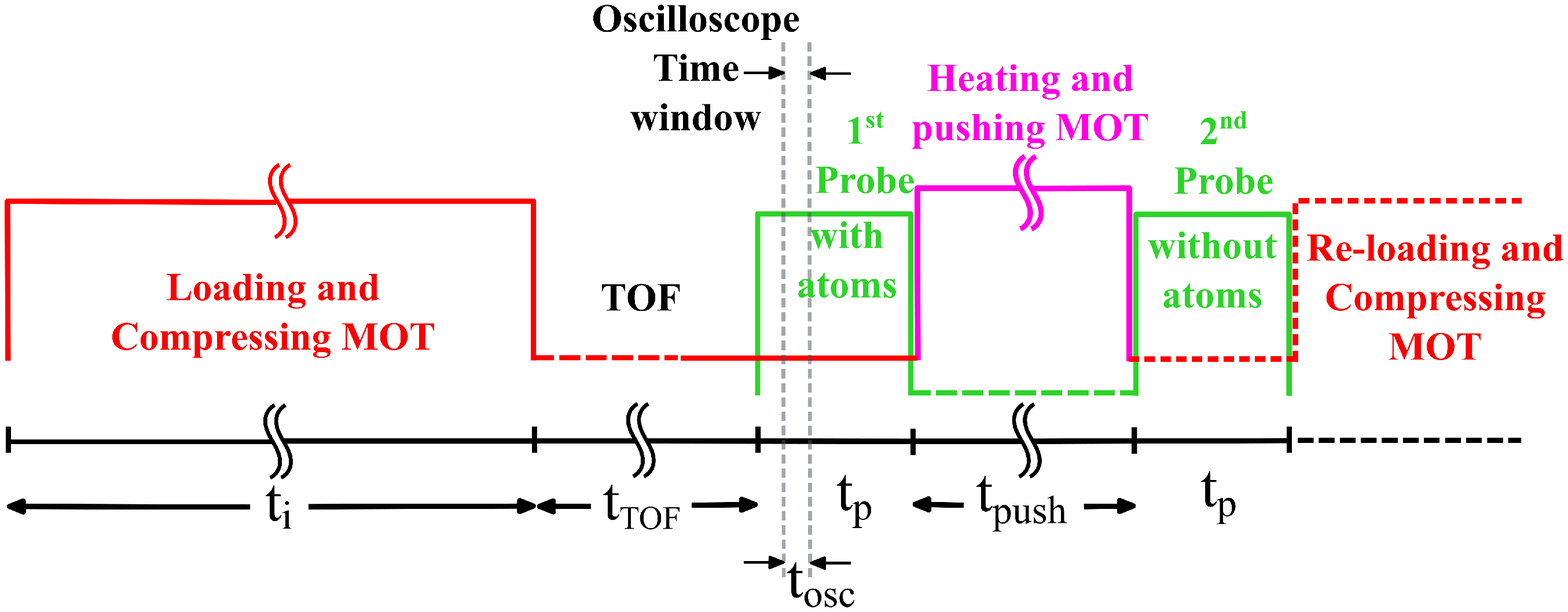}}
	\caption{(a) Schematic of our noise detection setup. PD: photodiode, Amp.: low noise amplifier. (b) Experimental time sequence. Typically $10^{10}$ atoms are loaded in the MOT and compressed during $t_\mathrm{i}$. Then the trapping system switches off and atoms are released. Two probe pulses are applied during $t_\mathrm{p} = 1.2$\,ms, after a time of flight of $t_\mathrm{TOF} = 4$\,ms. The first pulse provides the transmission through the atomic cloud, and the second one allows us to measure the incident intensity without atoms in order to calculate the normalized transmission for each cycle. The atoms are removed by applying the MOT beams at resonance during $t_\mathrm{push} = 6$\,ms between the two probe pulses. For the PSD measurements the time window of the oscilloscope is set $t_\mathrm{pause}=200$\,$\mu$s after the beginning of the first probe within $t_\mathrm{osc}=100$\,$\mu$s.}
\end{figure}

The experimental setup, based on measuring the intensity noise of a weak probe beam transmitted through a cloud of cold atoms, is depicted in Fig.\,\ref{fig:Cold_atoms_setup}.
The atomic cloud is obtained by loading a magneto-optical trap (MOT) with $^{85}$Rb atoms.
A compression is applied to increase the atomic density\,\cite{DePue_2000}.
The maximum number of atoms is $N \simeq 10^{10}$ with a temperature of about $100$\,$\mu$K and a cloud rms radius of $R \simeq 1$\,mm.

The probe beam is delivered by a distributed-feedback (DFB) laser. This laser, also used for the MOT beams, is amplified by a tapered amplifier. The laser frequency is locked using a master/slave configuration with an offset locking scheme\,\cite{Puentes_2012} and set close to the $F = 3 \rightarrow F' = 4$ hyperfine transition of the D2 resonance line of $^{85}$Rb. A double-pass acousto-optical modulator (AOM) is used to change the laser detuning $\delta = (\omega_\mathrm{L} - \omega_0)/\Gamma$ from this transition. The parameter $\omega_\mathrm{L}/2\pi = \nu_\mathrm{L}$ is the laser frequency, $\omega_0/2\pi$ the atomic transition frequency and $\Gamma/2\pi = 6.07$\,MHz the natural transition linewidth. The laser beam inside the atomic cloud is linearly polarized and its waist is about 300\,$\mu$m. The intensity is adjusted to have a saturation parameter lower than 0.1. The measurements are realized after a fixed $t_\mathrm{TOF} = 4$\,ms time of flight (TOF) and the laser beam path has been aligned to correspond to the centre of the atomic cloud after this TOF.

The laser beam after propagation through the atomic cloud is collected by a homemade transimpedance photodiode which has two outputs ports. The first port, corresponding to the DC output, is used to measure the probe transmission. The intensity noise is measured thanks to the AC output of the photodiode amplified by a low noise AC amplifier. The frequency response of the photodiode has been measured by illuminating it with a shot noise limited thermal light bulb. Its bandwidth ranges from about 10\,kHz to 10\,MHz. The power spectral density (PSD) of the detected signal is computed by an oscilloscope and the measured intensity noise PSD is finally normalized by the frequency response of the detection system.

The probe beam is applied during $t_\mathrm{p} = 1.2$\,ms but we fix the oscilloscope time window to $t_\mathrm{osc} = 100$\,$\mu$s, $t_\mathrm{pause} = 200$\,$\mu$s after the beginning of the probe pulse. To increase the signal to noise ratio, data are integrated over 100 cycles. We also record for each cycle the power of the probe beam without atoms $I_0$, which is needed for intensity to frequency noise conversion. This measurement is done by applying a second probe pulse after having removed all the atoms by shining the MOT beams at resonance during $t_\mathrm{push} = 6$\,ms. The time sequence is sketched in Fig.\,\ref{fig:Cold_atoms_TS}. The duration $t_\mathrm{i}$ includes the loading and compression stages. The optical thickness is varied by changing the total number of atoms through the MOT loading time.

\subsection{Transmission curve and frequency discriminator}\label{Par:FreqDiscr_Atoms}

\begin{figure}[h]
	\begin{center}
		\includegraphics[width=8cm]{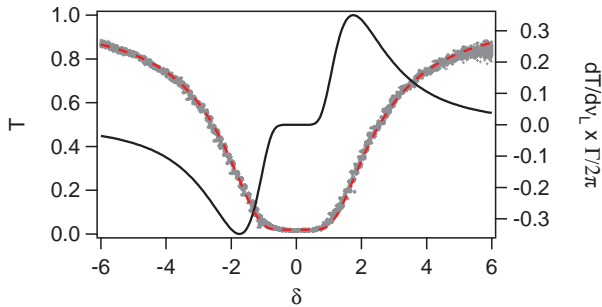}
		\caption{\label{Fig:Transmission_curve} Grey curve: experimental probe transmission as a function of the laser detuning. Red (dashed) curve: fit of the transmission curve, giving an on-resonant optical thickness of 19. Black curve: fit derivative used for the intensity to frequency noise conversion (right axis).}
	\end{center}
\end{figure}

Before noise measurements, we acquire the transmission curve by scanning the laser through the atomic transition. A typical transmission curve is plotted in Fig.\,\ref{Fig:Transmission_curve}. The on-resonance optical thickness $b_0$ is extracted by fitting the data by the expected transmission,
\begin{eqnarray}
T &\equiv& \frac{I_\mathrm{a}}{I_0} = e^{-b(\delta)},\label{Eq:T}
\end{eqnarray}
with
\begin{eqnarray}
b(\delta) &=& \frac{b_0}{1+4\delta^2}.\label{Eq:b}
\end{eqnarray}
The parameter $I_\mathrm{a}$ corresponds to the power measured with atoms while $I_0$ is measured without atoms.

The dependance of the transmission with the laser frequency allows us to use it as a frequency discriminator. The relation between the PSD $S_{T}$ in [Hz$^{-1}$] of the transmitted intensity normalized by the intensity without atoms, $T = I_\mathrm{a}/I_0$, and the laser frequency noise PSD (FNPSD) $S_{\nu_\mathrm{L}}$ in [Hz$^2$/Hz] is given by the following equation,
\begin{eqnarray}
S_{T} &=& \left(\frac{\mathrm{d}T}{\mathrm{d}\nu_\mathrm{L}}\right)^2 S_{\nu_\mathrm{L}}\label{eq:Conversion_atoms}\\
&=& D^2 S_{\nu_\mathrm{L}},
\end{eqnarray}
valid when the transmission curve can be locally approximated by a line whose slope is equal to $D$. As we will see in Sec.\,\ref{SubSec:Modeling}, this is a good approximation for low Fourier frequencies. Using Eqs.\,(\ref{Eq:T}) and (\ref{Eq:b}), one finds:
\begin{equation}
D = \frac{\mathrm{d}T}{\mathrm{d}\nu_{\mathrm{L}}} = \frac{16\pi \delta b_0 }{\Gamma\left(1+4\delta^2\right)^2} e^{-\frac{b_0}{1+4\delta^2}}.\label{eq:dT/ddelta}
\end{equation}
The discriminator slope $D$ depends on the optical thickness $b_0$ as well as the laser detuning\,$\delta$. A typical curve is plotted in Fig.\,\ref{Fig:Transmission_curve} for an optical thickness of 19. The best conversion is obtained when $|D|$ is maximum, corresponding to $|\delta| \simeq 1.8$ for $b_0 = 19$.

\subsection{Experimental results}

We have measured the transmitted intensity noise PSD for three different on-resonance optical thicknesses: $b_0 = 6.5$, 19 and 51.5. For each $b_0$, the laser detuning is adjusted, thanks to the double-pass AOM, to be at the maximum of the discriminator slope on the blue side of the atomic transition. We have first checked that the detection background and the intrinsic laser intensity noise, measured without atoms, are well below the intensity noise PSD measured with atoms. The PSD $S_T$ is then converted to FNPSD using Eq.\,(\ref{eq:Conversion_atoms}).

\begin{figure}[h]
\begin{center}
\includegraphics[width=8cm]{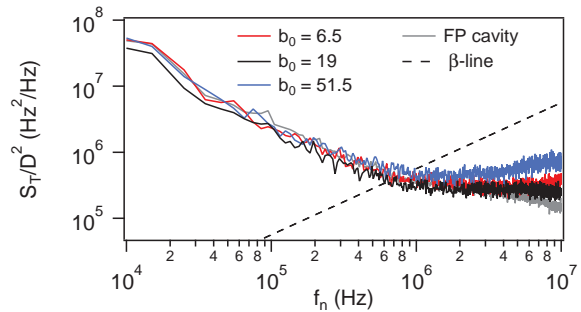}
\caption{\label{Fig:Freq_noise_Atoms} Laser transmission noise PSD $S_T$ divided by the square of the discriminator slope $D^2$, measured using a cold atomic cloud as a frequency discriminator (from the black curve to the top: $b_0 = 19$, 6.5 and 51.5). For low Fourier frequencies, Eq.\,(\ref{eq:Conversion_atoms}) is valid and the curves thus correspond to the laser frequency noise PSD (FNPSD). The FNPSD measured with a Fabry-Perot cavity as a frequency discriminator is plotted in grey (see appendix\,\ref{App:FP} for more details). The dashed line corresponds to the $\beta$-separation line used to estimate the laser linewidth.}
\end{center}
\end{figure}

The results are plotted in Fig.\,\ref{Fig:Freq_noise_Atoms} for three optical thicknesses. Whereas all the PSDs are consistent at low frequencies, typically below 1\,MHz, some differences appear at higher frequencies.

\section{Noise spectroscopy at low frequencies: laser spectral properties}\label{Sec:LowFreq}

For low Fourier frequencies, typically for $f_\mathrm{n}$ below 1\,MHz, Eq.\,(\ref{eq:Conversion_atoms}) is valid and the curves $S_{T}/D^2$ plotted in Fig.\,\ref{Fig:Freq_noise_Atoms} thus correspond to the laser FNPSD $S_{\nu_\mathrm{L}}$. This is also confirmed by comparing these curves to the one obtained with a Fabry-Perot (FP) cavity used as frequency discriminator (see appendix\,\ref{App:FP} for more details).

The measurements of the laser FNPSD can be used to recover the laser linewidth. The first method is to use the formula which links the FNPSD and the optical spectrum\,\cite{Elliott_1982}:
\begin{eqnarray}
&S_E(\nu)& =  2 \int_{-\infty}^{\infty} e^{-2i \pi \nu \tau} \bigg[E_0^2 e^{2i \pi \nu_\mathrm{L} \tau} \nonumber \\
&& \mathrm{exp} \left(-2 \int_{0}^{\infty} S_{\nu_\mathrm{L}}(f_\mathrm{n}) \frac{\sin^2(\pi f_\mathrm{n} \tau)}{f_\mathrm{n}^2} df_\mathrm{n} \right)\bigg]d\tau.
 \label{eq:SE_Snu}
\end{eqnarray}
However, most of the time, there is no analytical solution for this equation and one needs to perform a tedious numerical integration. As an alternative method, the laser linewidth can be estimated using a simple geometrical approach, based on the so-called $\beta$-separation line. Since the first theoretical study\,\cite{DiDomenico_2010}, this approach has been applied in many experimental setups\,\cite{Llopis_2011,Bucalovic_2012,Dinesan_2015,Ricciardi_2015} and further refined\,\cite{Zhou_2015,Zhou2_2015}.

The $\beta$-separation line is defined as\,\cite{DiDomenico_2010}:
\begin{equation}
S_\beta(f_\mathrm{n}) = \frac{8 \text{ln}(2)}{\pi^2} f_\mathrm{n}.
\label{eq:S_beta}
\end{equation}
This line, plotted in Fig.\,\ref{Fig:Freq_noise_Atoms} and in Fig.\,\ref{fig:freq_noise_FP}, separates the noise spectrum in two regions. For $S_\beta(f_\mathrm{n}) < S_{\nu_\mathrm{L}}(f_\mathrm{n})$, frequency noise contributes to the central Gaussian part of the laser line shape and thus to the linewidth. For $S_\beta(f_\mathrm{n}) > S_{\nu_\mathrm{L}}(f_\mathrm{n})$, frequency noise contributes to the Lorentzian wings of the line shape and does not significantly affect the linewidth. This method approximates the laser linewidth\,$\Delta\nu_\mathrm{L}$, corresponding to the FWHM of the central part of the line shape, by
\begin{equation}
\Delta\nu_\mathrm{L} = \sqrt{8 \text{ln}(2) A},
\label{eq:linewidth_beta}
\end{equation}
with $A$ the area below $S_{\nu_\mathrm{L}}$ in the frequency range where $S_{\nu_\mathrm{L}}$ is above the $\beta$-separation line, i.e.:
\begin{equation}
A = \int_{1/T_\mathrm{obs}}^{+ \infty} H \big[S_{\nu_\mathrm{L}}(f_\mathrm{n}) - S_\beta(f_\mathrm{n}) \big] S_{\nu_\mathrm{L}}(f_\mathrm{n}) \mathrm{d}f_\mathrm{n},
\label{eq:A}
\end{equation}
where $T_\mathrm{obs}$ is the observation time, and $H$ is the Heaviside step function.

The values obtained with this approach are listed in Table\,\ref{Tab:Linewidth} together with the values obtained with the Fabry-Perot cavity. The uncertainties obtained with the cold atomic cloud take into account the statistical uncertainty (standard deviation of the linewidth measurements obtained in similar conditions) as well as the estimation of the maximum error due to the $\beta$-separation line approach\,\cite{Zhou_2015}. The laser linewidth has also been measured using the beat-note technique (see appendix\,\ref{App:FP}). All these results are compatible, validating the fact that the noise measured with the cold atomic sample at low Fourier frequencies corresponds to the intrinsic laser frequency noise.

\begin{table}[h]
	\centering
	\begin{tabular}{m{4cm} | m{3cm}}
				\hline \hline
\centering Experimental technique & \centering Linewidth
\tabularnewline \hline
Cold atomic cloud &
\tabularnewline
     $\qquad b_0 = 6.5$ & \centering $3.7 \pm 0.5$\,MHz
\tabularnewline
     $\qquad b_0 = 19$ & \centering $3.3 \pm 0.5$\,MHz
\tabularnewline
     $\qquad b_0 = 51.5$ & \centering $3.7 \pm 0.5$\,MHz
\tabularnewline
FP cavity &  \centering $3.4 \pm 0.4$ MHz
\tabularnewline
Beat-note  &  \centering $3.0 \pm 0.2$ MHz
\tabularnewline
		\hline	\hline
	\end{tabular}
\caption{DFB laser linewidth obtained with different techniques.}
	\label{Tab:Linewidth}
\end{table}

\section{Noise spectroscopy at high frequencies: atomic cloud properties}\label{Sec:HighFreq}

\subsection{Experimental results}

We can see in Fig.\,\ref{Fig:Freq_noise_Atoms} that the PSDs differ at high frequencies, with in particular the appearance of a small ``bump''. However, these curves becomes limited by the noise floor of the photodiode for frequencies higher than 1\,MHz. To overcome this problem, the photodiode and the amplifier have been replaced by a new photodiode, with a high cutoff frequency of 240\,MHz and with a lower noise floor.

\begin{figure}[h]
\begin{center}
\includegraphics[width=8cm]{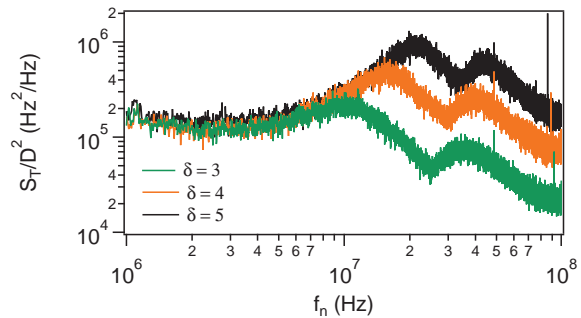}
\caption{\label{Fig:f_PSD_Bump} Zoom at high Fourier frequencies of the laser transmission noise PSD $S_T$ divided by the square of the discriminator slope $D^2$ using a cold atomic cloud with an optical thickness of $b_0 = 19$ and for three different laser detunings (lower curve: $\delta = 3$; curve in the middle: $\delta = 4$; upper curve: $\delta = 5$).}
\end{center}
\end{figure}

Typical FNPSDs, obtained with this low noise photodiode, are zoomed at high frequencies in Fig.\,\ref{Fig:f_PSD_Bump}. The three curves have been measured with the same optical thickness $b_0 = 19$ but for three different laser detunings. We clearly see the appearance of bumps whose frequency positions depend on the laser detuning. These positions also depend on the optical thickness as shown in Fig.\,\ref{Fig:Bump_position}, where the frequency position of the first and second bump is plotted as a function of the laser detuning and for the three previous optical thicknesses.

\subsection{Modeling}\label{SubSec:Modeling}

To understand what happens at high frequencies, we model the laser frequency noise as a carrier at frequency $\nu_\mathrm{L}$ with two weak sidebands at $\nu_\mathrm{L} \pm f_\mathrm{n}$ as done in Ref.\,\cite{Bahoura_2001}. It corresponds to a phase modulation where the laser field can be written as follows:
\begin{eqnarray}
E &=& E_0 e^{i[2\pi  \nu_\mathrm{L} t + B_\phi \sin (2\pi f_\mathrm{n}t)]},\\
&\simeq& E_0\left[e^{i2\pi\nu_\mathrm{L} t} + \frac{B_\phi}{2}e^{i2\pi(\nu_\mathrm{L}+f_\mathrm{n}) t} - \frac{B_\phi}{2}e^{i2\pi(\nu_\mathrm{L}-f_\mathrm{n}) t} \right].\nonumber\\
\end{eqnarray}
The parameter $B_\phi$ corresponds to the phase modulation depth. The corresponding amplitude of the frequency noise at frequency $f_\mathrm{n}$ is $B_\phi f_\mathrm{n}$.

When the laser goes through the atoms, the carrier and the two sidebands experience different transmissions $\sqrt{T}$ and phase shifts $\phi$:
\begin{eqnarray}
\sqrt{T(\delta)} &=& e^{-\frac{b_0}{2(1+4\delta^2 )}},\\
\phi(\delta) &=& -\frac{b_0 \delta}{1+4\delta^2}.
\end{eqnarray}
The laser intensity transmission as a function of time $t$ becomes
\begin{equation}
\frac{I_\mathrm{a}}{I_0} = T_0 \left[ 1+\frac{B_\phi C_\phi}{\sqrt{T_0}} \cos{(2\pi f_\mathrm{n}t + \psi)}\right],
\end{equation}
where $\psi$ is a phase shift and
\begin{equation}
C_\phi = \sqrt{T_1 + T_2-2\sqrt{T_1T_2} \cos(2\phi_0 - \phi_1 - \phi_2)},\label{eq:C}
\end{equation}
where $T_0$ and $\phi_0$ are the intensity transmission and phase shift induced by the atoms on the carrier and $T_1$, $T_2$, $\phi_1$, $\phi_2$ the intensity transmission and phase shifts induced on the two sidebands. The theoretical frequency to intensity noise conversion is thus given by:
\begin{equation}
S_{T,\mathrm{th}} = \left(\frac{\delta I_\mathrm{a}}{I_0}\right)^2 = T_0\frac{C_\phi^2}{f_\mathrm{n}^2} S_{\nu_\mathrm{L}}.\label{eq:Conversion_atoms_model}
\end{equation}
We can show that we recover the conversion given by Eqs.\,(\ref{eq:Conversion_atoms}) and (\ref{eq:dT/ddelta}) for $f_\mathrm{n} \ll \Gamma$ and $f_\mathrm{n} \ll \delta \Gamma$ respectively.

The same approach can be used if one deals with laser amplitude noise instead of frequency noise. In this case, the laser field can be written as follows:
\begin{eqnarray}
E &=& E_0\left[1+B_E \sin (2\pi f_\mathrm{n}t) e^{i2\pi  \nu_\mathrm{L} t}\right],\\
&=& E_0\left[e^{i2\pi\nu_\mathrm{L} t} - i\frac{B_E}{2}e^{i2\pi(\nu_\mathrm{L}+f_\mathrm{n}) t} + i \frac{B_E}{2}e^{i2\pi(\nu_\mathrm{L}-f_\mathrm{n}) t} \right], \nonumber\\
\end{eqnarray}
with $B_E$ the amplitude modulation depth. The laser intensity transmission is calculated as previously and one gets:
\begin{equation}
\frac{I_\mathrm{a}}{I_0} = T_0 \left[ 1+\frac{B_E C_E}{\sqrt{T_0}} \cos{(2\pi f_\mathrm{n}t + \psi)}\right],
\end{equation}
with
\begin{equation}
C_E = \sqrt{T_1 + T_2 + 2\sqrt{T_1T_2} \cos(2\phi_0 - \phi_1 - \phi_2)}.\label{eq:C_E}
\end{equation}
The normalized intensity PSD of the transmitted beam is finally given by:
\begin{equation}
S_{T,\mathrm{th}} = \left(\frac{\delta I_\mathrm{a}}{I_0}\right)^2 = T_0 C_E^2 S_E,
\end{equation}
where $S_E$ corresponds to the normalized laser amplitude noise PSD.

In either case, the shape of $S_T$ depends on the Fourier frequency through the phase shifts $\phi_1$ and $\phi_2$ induced by the atoms. We can also see that the two types of noises give the same typical equations with only a change of sign in Eqs.\,(\ref{eq:C}) and (\ref{eq:C_E}). This change of sign is then responsible for the shape dependance of the PSD $S_T$ on the type of laser noise.

\subsection{Comparison between experimental and modeling results}\label{SubSec:Comparison}

The previous model is used to calculate the expected noise of the transmitted intensity. We assume a white frequency noise for frequencies higher than 1\,MHz. Its value, extracted from the measured PSD at $f_\mathrm{n}= 1$\,MHz, is set to $(B_\phi f_\mathrm{n})^2 \simeq S_{\nu_\mathrm{L}} \simeq 10^{5}$\,Hz$^2/$Hz.
This is injected in Eq.\,(\ref{eq:Conversion_atoms_model}) to calculate the expected transmission noise PSD $S_{T,\mathrm{th}}$. We then divide it by the discriminator slope given by Eq.\,(\ref{eq:dT/ddelta}) in order to compare it to the measurements.

\begin{figure}[h]
\begin{center}
\includegraphics[width=8cm]{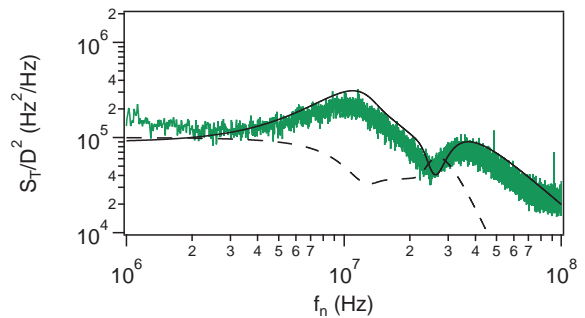}
\caption{\label{Fig:Freq_noise_Exp_Simu} Solid green line: laser transmission noise PSD $S_T$ divided by the square of the discriminator slope $D^2$ using a cold atomic cloud with an optical thickness of $b_0 = 19$ and a laser detuning of $\delta = 3$. Solid black (thin) line: $S_{T,\mathrm{th}}/D^2$ calculated using Eqs.\,(\ref{eq:C}) and (\ref{eq:Conversion_atoms_model}) assuming a white frequency noise. Dashed line: $S_{T,\mathrm{th}}/D^2$ calculated using Eq.\,(\ref{eq:C_E}) assuming a white amplitude noise.}
\end{center}
\end{figure}

Typical calculated and measured PSD are compared in Fig.\,\ref{Fig:Freq_noise_Exp_Simu}. The optical thickness is $b_0 = 19$ and the laser detuning is $\delta = 3$. We see a good overlap between the measured and the calculated PSD, without any free parameter. In particular, the model predicts the existence of two bumps whose frequency positions correspond to the ones experimentally observed. These bumps are intrinsically related to the fact that we deal with frequency noise. On the contrary, if we do the calculations assuming an incident laser with amplitude noise, one sees the appearance of dips instead of bumps as shown in Fig.\,\ref{Fig:Freq_noise_Exp_Simu}. The analysis of the noise spectra at high Fourier frequencies, with the presence of bumps or dips, thus allows to extract the nature of the laser noise.

\begin{figure}[h]
\begin{center}
\includegraphics[width=8cm]{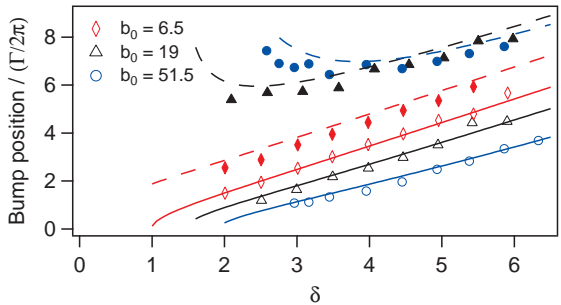}
\caption{\label{Fig:Bump_position} Points: experimental frequency position of the bumps (empty icon: first bump, filled icon: second bump) observed in the FNPSD, obtained with the cold atomic cloud, as a function of the laser detuning. Solid line: calculated frequency position of the first bump. Dashed line: calculated frequency position of the second bump assuming frequency noise.}
\end{center}
\end{figure}

Finally, we have compared the measured and the calculated bump positions, corresponding to the frequency position of the local maxima, as a function of the laser detuning and for the three different optical thicknesses. The results are plotted in Fig.\,\ref{Fig:Bump_position}, assuming laser frequency noise for the calculations. We obtain a very good agreement between measurements and calculations, validating the model used to understand the frequency to intensity noise conversion. We can also notice that the frequency difference between both bump positions remains constant, at least for sufficiently high laser detuning, and that this difference roughly corresponds to the frequency range where the transmission curve is close to zero. Both bumps can thus be interpreted as the signature of the beat-note between the carrier and the phase shifted and attenuated sideband that goes from one side of the transmission curve to the other.

The frequency positions of the bumps in the FNPSD as well as the difference between both bump positions depend on the laser detuning from the atomic transition and on the optical thickness. Their measurements thus allow extracting information on the discriminator medium itself. One could imagine to extract the two last quantities by directly measuring one FNPSD instead of measuring the entire transmission curve by scanning the laser frequency around the atomic transition.

\section{Conclusion}

In this paper, we have studied the intensity noise on a laser beam transmitted through a cold atomic cloud. In this forward configuration, we have observed the conversion of the intrinsic laser frequency noise to intensity noise, the atomic transition playing the role of a frequency discriminator whose slope is adjustable through the optical thickness. While we recover the same FNPSD at low Fourier frequencies using a Fabry-Perot cavity, some differences appear at higher Fourier frequencies and one needs to go beyond the linear response approximation of the discriminator given in Eq.\,(\ref{eq:Conversion_atoms}). We have shown that a simple model in which the frequency noise is modeled as a carrier with two sidebands and the atomic cloud as a medium with an index of refraction is sufficient to describe the observations.

The measurement of one single spectrum of the intensity noise allows us to extract much information. The conversion of the laser frequency noise to intensity noise can be used to characterize the laser noise. However one can also extract the nature of the laser noise or some important characteristics of the atomic sample such as its optical thickness. Usually perceived as a drawback, frequency to intensity noise conversion can clearly be seen as an important source of information.

Finally, in this forward direction, the conversion of the intrinsic laser frequency noise to intensity noise is usually an important source of noise. An accurate understanding of this effect is thus of crucial importance. With this conversion now well characterized, intensity noise measurements could possibly be used to extract some signatures of more involved phenomena, such as the observation of the influence of atomic motion and quantum optical properties\,\cite{Peyronel_2012, Lambrecht_1996, Ries_2003, Lezama_2015}.

\acknowledgments

We thank Gabriel H\'etet for the very first measurements on this setup. A. Eloy acknowledges the support of the DGA.


\appendix

\section{Intrinsic laser noise characterization}\label{App:FP}

The aim of this appendix is to present the characterization of our laser intrinsic noise using standard techniques. The corresponding results serve as a benchmark for FNPSD measured with cold atoms.

The laser is the one used as the probe beam in our cold atoms experiment. It is delivered by a DFB laser amplified by a tapered amplifier, and frequency locked close to the $F = 3 \rightarrow F' = 4$ hyperfine transition of the D$_2$ line of $^{85}$Rb. The laser spectral properties can be described through the optical spectrum and the corresponding linewidth or through the frequency noise power spectral density (FNPSD). These two complementary approaches are presented in this appendix.
\\

\subsection{Line shape and linewidth} \label{Sec:Linewidth}

\begin{figure}[h]
	\centering
	\includegraphics[width=0.9\linewidth]{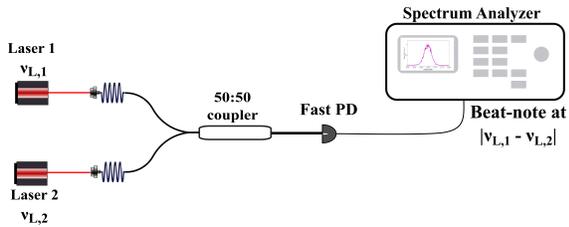}
	\caption{Setup for the beat-note measurement. Two laser beams are injected in a $50:50$ coupler. Both lasers are independently frequency locked, and the frequency difference between them is typically 1\,GHz. The interference signal is collected by a fast photodiode (PD) and then analyzed by a spectrum analyzer.}
	\label{fig:BN_setup}
\end{figure}

The optical spectrum corresponds to the power spectral density (PSD) of the laser electric field $S_E(\nu)$. It is measured using a beat-note technique sketched in Fig.\,\ref{fig:BN_setup}. The beat-note signal is obtained using two different lasers. The first one is the DFB laser we want to characterize. For the second laser we have used a DL Pro commercial external-cavity diode laser (ECDL) from TOPTICA, whose linewidth is specified to be lower than 500\,kHz\,\footnote{This specification is given for a free-running laser with a measurement time of 5\,$\mu$s}. The lasers are independently frequency locked and the frequency difference between them is typically 1\,GHz. The two lasers are injected into a $50:50$ fiber coupler. The beat-note signal is then detected by a $9.5$\,GHz bandwidth photodiode and its power spectral density is measured by a spectrum analyzer.

\begin{figure}[h]
	\centering
	\includegraphics[width=8cm]{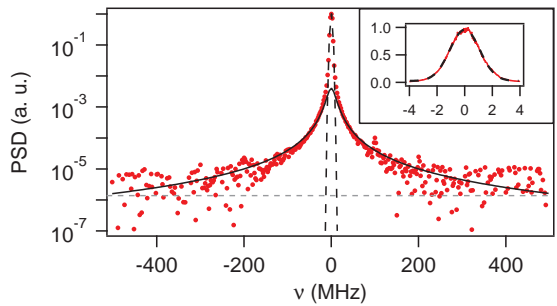}
	\caption{Beat-note signal PSD between the DFB laser and the TOPTICA laser. The PSD has been averaged 100 times with a sweep time of 0.5\,s. The center of the spectrum has been shifted to the origin. Since the TOPTICA laser has a much smaller linewidth than the DFB laser one, it can be treated as a reference laser and the PSD mainly corresponds to the optical spectrum of the DFB laser. The central part can be fitted by a Gaussian (dashed line) with a FWHM $\Delta\nu_\mathrm{BN} \simeq 3$\,MHz, and the wings are well fitted by a Lorentzian (solid line). The horizontal grey dashed line corresponds to the typical noise floor. Inset: zoom on the Gaussian part of the optical spectrum. Red curve: beat-note signal PSD. Dashed curve: Gaussian fit.}
	\label{fig:Optical Spectrum}\end{figure}

The beat-note PSD is plotted in Fig.\,\ref{fig:Optical Spectrum}. It corresponds to the convolution of the optical spectrum of the two lasers, but since the linewidth of the TOPTICA laser is much smaller than the DFB laser one, it is mainly dominated by the DFB optical spectrum. It contains a central part at the frequency difference of the two lasers, which can be fitted by a Gaussian, superposed on large wings, which can be fitted by a Lorentzian. The linewidth, given by the full width at half-maximum (FWHM) of the gaussian part, is $\Delta\nu_\mathrm{BN} \simeq 3$\,MHz. The Lorentzian part has a FWHM of 20\,MHz and an amplitude typically one thousand times smaller than the Gaussian part.

Due to the convolution of the two laser Gaussian optical spectra, the square of the beat-note FWHM is the quadratic sum of the two laser linewidths $\Delta \nu_\mathrm{L}$:
\begin{equation}
	\Delta \nu_\mathrm{BN}^2 = \Delta \nu_{\mathrm{L},1}^2 + \Delta \nu_{\mathrm{L},2}^2.
	\label{eq:BN_width}
\end{equation}
To deduce each laser linewidth, three different beat-note signals from three different lasers are needed. The third one is a homemade, etalon-based ECDL with an intermediate linewidth~\cite{Baillard_2006}. The results are summarized in Tab.\,\ref{tbl:tbl_linewidth}. The uncertainties are given at $1\sigma$ and have been obtained with a statistical analysis of different beat-note signals recorded in the same conditions.

\begin{table}[h]
	\centering
	\begin{tabular}{m{3cm} | m{4cm}}
				\hline \hline
\centering Laser & \centering Linewidth
\tabularnewline \hline
		\centering TOPTICA  & \centering $0.2\,(+1.5/-0.2)$ MHz
\tabularnewline
		\centering homemade ECDL &  \centering $1.1\,(\pm 0.3)$ MHz
\tabularnewline
		\centering DFB &  \centering $3.0\,(\pm 0.2)$ MHz
\tabularnewline
		\hline	\hline
	\end{tabular}
\caption{Laser linewidth measured using the beat-note setup, with lasers independently frequency locked. The uncertainties are given at $1\sigma$.}
	\label{tbl:tbl_linewidth}
\end{table}

\subsection{Frequency noise PSD} \label{Sec:Freq_noise}

Measurements of the laser optical spectrum, and the corresponding linewidth, and the laser FNPSD are complementary. But, while the first one is convenient to quickly compare different types of lasers, the second one gives a much more complete knowledge of the laser spectral properties.

One of the most common frequency discriminator, used to measure laser FNPSD $S_{\nu_\mathrm{L}}(f_\mathrm{n})$, is the Fabry-Perot (FP) cavity. Its transmission depends on the light frequency which allows to convert frequency noise into intensity noise. For a laser linewidth smaller than the cavity linewidth $\Delta\nu_\mathrm{c}$, the conversion between the FNPSD and the normalized intensity noise PSD (INPSD) of the transmitted beam is
\begin{equation}
S_{I_\mathrm{n}} = \left(\frac{\mathrm{d}T_\mathrm{c}}{\mathrm{d} \nu_\mathrm{L}}\right)^2 S_{\nu_\mathrm{L}},
\label{eq:conversion_FP}
\end{equation}
with $T_\mathrm{c}$ the cavity transmission, $\nu_\mathrm{L}$ the central laser frequency and $S_{I_\mathrm{n}}$ the PSD of the transmitted intensity normalized by the incident intensity. The parameter $D = \mathrm{d}T_\mathrm{c}/\mathrm{d} \nu_\mathrm{L}$ is the discriminator slope. The optimum conversion is obtained when the laser is tuned to the half maximum of the cavity resonance, where $D$ reaches its maximum $1/\Delta\nu_\mathrm{c}$.

\begin{figure}[h]
	\centering
	\includegraphics[width=0.9\linewidth]{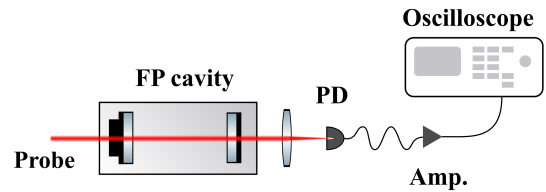}
	\caption{Schematic of frequency noise PSD measurement using a Fabry-Perot (FP) cavity as a frequency discriminator. PD: photodiode, Amp.: low noise amplifier.}
	\label{fig:FP_setup}
\end{figure}

The experimental setup is shown in Fig.\,\ref{fig:FP_setup}. We use a free-running confocal FP cavity with a length $L_\mathrm{c} \simeq 10$\,cm.
The corresponding free spectral range is $\Delta \nu_\mathrm{FSR} = c/4L_\mathrm{c} \simeq 750$\,MHz and the cavity linewidth is $\Delta \nu_\mathrm{c} \simeq 20 $\,MHz. A piezoelectric transducer is placed on one of the mirrors, allowing us to adjust the cavity-to-laser detuning to the half of the cavity resonance. The laser transmitted by the cavity is detected by the same homemade transimpedance photodiode used for the first measurements with the cold atomic cloud. The PSD of the detected signal is recorded by an oscilloscope computing its power spectrum and normalized by the frequency response of the detection system. It is finally converted to FNPSD using Eq.\,(\ref{eq:conversion_FP}).

\begin{figure}
	\centering
	\includegraphics[width=8cm]{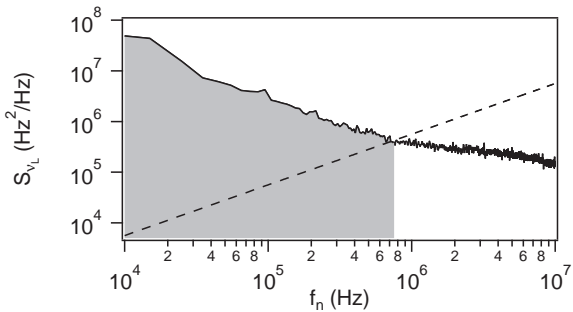}
	\caption{Frequency noise PSD for the DFB laser. The dashed line corresponds to the $\beta$-separation line used to estimate the laser linewidth. Grey area: contribution to the DFB laser linewidth. See section\,\ref{Sec:LowFreq} for further details.}
	\label{fig:freq_noise_FP}
\end{figure}

The FNPSD for the DFB laser is depicted in Fig.\,\ref{fig:freq_noise_FP}. We have checked that the PSD is not limited by the detection background or by the intrinsic laser intensity noise.
The FNPSD essentially decreases as $1/f_\mathrm{n}$ up to 1\,MHz, corresponding to a flicker noise.

As done in section\,\ref{Sec:LowFreq}, we can extract the laser linewidth using the $\beta-$separation line approach. We obtain a linewidth of $3.4\pm 0.4$\,MHz for the DFB laser. The 10$\,\%$ relative uncertainty takes into account the maximum typical error introduced by the $\beta$-line approach\,\cite{Zhou_2015}, which is an approximate method to estimate the laser linewidth. The laser linewidth is compatible with the one obtained from the beat-note measurement.

\end{document}